\documentclass[%
preprint,
showpacs,
amsmath,amssymb,
aps,
]{revtex4-1}

\usepackage{graphicx}
\usepackage{color}
\usepackage{dcolumn}
\usepackage{bm}
\usepackage{setspace}

\begin{document}

\preprint{}

\title{
Phase transition to two-peaks phase in an information cascade voting experiment
}

\author{Shintaro Mori}
 \email{mori@sci.kitasato-u.ac.jp}
 \affiliation{Department of Physics, Kitasato University \\
Kitasato 1-15-1, Sagamihara,  Kanagawa 252-0373, JAPAN
}

\author{Masato Hisakado}
\affiliation{
Standard and Poor's 
\\
Marunouchi 1-6-5, Chiyoda-ku, Tokyo 100-0005, JAPAN
}%

\author{Taiki Takahashi}
\affiliation{%
Department of Behavioral Science, Hokkaido University
\\
N.10,W.07, Kita-ku, Sapporo, Hokkaido 060-0810, JAPAN
}%

\date{\today}

\begin{abstract}
Observational learning is an important information aggregation mechanism. 
However, it occasionally leads to a state in which an entire population chooses 
a sub-optimal option. When this occurs and whether it is a phase transition remain 
unanswered. 
To address these questions, we performed a voting experiment in which subjects answered 
a two-choice quiz sequentially with and without information about the prior subjects' choices. 
The subjects who could copy others are called herders. 
We obtained a microscopic rule regarding how herders copy others. Varying the ratio 
of herders led to qualitative changes in the macroscopic behavior of about 50 subjects in the experiment. 
If the ratio is small, the sequence of choices rapidly converges 
to the correct one.  As the ratio approaches 100\%, convergence becomes extremely slow and 
information aggregation almost terminates.
A simulation study of a stochastic model for $10^{6}$ subjects based on 
the herder's microscopic rule 
showed a phase transition to the two-peaks phase, where the convergence 
completely terminates  as the ratio exceeds some critical value.
\end{abstract}

\maketitle


\section{\label{sec:intro}Introduction}

 The tendency to imitate others is one of the basic instinct of human.
 \color{black} People effectively and inadvertently act as filters to provide the information that is most useful for an observer. 
 Imitation and copying is a highly adaptive means of gaining knowledge
 \cite{Ren:2010}
\color{black}.
 It presumably results from an evolutionary adaptation that promoted survival over thousands of generations.
 It allows individuals to exploit the hard-won information of others \cite{Ren:2011,Ren:2010}.
 However, imitating or copying others has disadvantages. 
 The acquired information might be 
 outdated or misleading \cite{Ren:2011,Lor:2011}. 
 Copying wrong information might lead to  herding, where an  
 entire population makes a wrong decision. This is referred to  
 as information cascade or rational herding \cite{Bik:1992,Dev:1996,And:1997,Lee:1993}. 
 Unfortunately, because imitation is a basic instinct and because it is
 economically rational to copy others, humans \color{black} might not be able to \color{black} evade such a 
 catastrophic situation \cite{Bik:1992}.
 \color{black} 
 Social influences have many forms, including  imitation, conformity and  obedience \cite{Kel:1958}. 
 Recent studies in cognitive neuroscience suggest that some type of imitation occurs automatically 
 via the actions of ``mirror neuron" systems \cite{Hey:2011}.
 \color{black}
 
 In the field of social psychology, many studies have focused on how humans use social  
 information at the microscopic level \cite{Ren:2011,Gri:2006}.   
 In the field of finance and economics, 
 it is now widely believed that investors are influenced by  the decisions of others
 and that this influence is a first-order effect \cite{Bik:1992,Dev:1996}.
 We now have a number of interesting models of rational herding based on simple,
 straightforward, 
 and convincing intuition \cite{Bik:1992,Kir:1993,Dev:1996,Lux:1995,Con:2000,Cur:2006,Mor:2010,Gal:2008,Gon:2011}.
 Empirical financial research has focused on macroscopic data  primarily because such
 data is easily available. 
\color{black}
 Micro-macro aspects of information cascade have been studied in \cite{Goe:2007}. It was concluded that 
 the information cascade is fragile and self-correcting. Even if the population makes 
 a wrong decision at a point in the choice sequence, it will eventually turn to a 
 correct choice. The analysis was based on a stochastic model, and the asymptotic behavior
 of the empirical choice sequences was not studied in detail. In order to study the nature of  information cascade and,
 furthermore the possibility of phase transition, 
 it is necessary to connect the micro-macro aspects without 
 depending on a model assumption as far as possible. 
\color{black}
 However, thus far, no empirical work has directly connected 
 the microscopic and macroscopic aspects of information cascade and herding.

  Two types of phase transitions have been predicted in a two-choice voting model 
  depending on the strength of conformity of people \cite{His:2010,His:2011}. 
 We set two types of individuals: herders and independents. The voting of independents is based on their 
 fundamental values, while the voting of herders is  based on the number of previous votes.
 If the herders are analog herders and they vote for each choice with probabilities that are proportional
 to the choices' votes, there occurs a transition between the super and normal diffusion phases \cite{His:2010}.
 If the independents are the majority of voters, the voting rate converges at the same rate as 
 in a binomial distribution, which is called the normal diffusion phase.
 As the proportion of herders is over 50\%, the voting rate converges more slowwly than in a binomial distribution: 
 this is called the super-diffusion phase. However, the presence of herders does not affect the 
 accuracy of the majority's choice. If the independents vote for the correct choice rather than for
 the wrong one, the majority of voters always choose the correct choice.
 The probability distribution of the voting rate has only one peak, and these
 two phases are collectively referred to as the one-peak phase. 
 In the digital herder case, where herders always choose the choice with the majority of previous votes,
 the majority's choice does not necessarily coincide with the correct choice, even if the independents vote 
 for the correct choice rather than for the wrong one.  
 When the fraction of herders increases, there occurs a phase transition, beyond which a state where
 most voters choose the correct choice coexists with one where most of them choose 
 the wrong one \cite{His:2011}.
 If the fraction of herders is below the threshold value, most voters 
 choose the correct choice and the system  is  in the one-peak phase.
 If the fraction is above the threshold value, the distribution of the voting rate has two peaks€€
 corresponding to the two coexisting states: this phase is called the two-peaks phase.
 We call the phase transition between the one-peak and two-peaks phases the information cascade 
 transition.

 In this paper, we have adopted an experimental and theoretical approach to the study of 
 the phase transitions \color{black} in information cascade \color{black}.
 The organization of the paper is as follows. 
 We explain the experimental design and procedure in section \ref{sec:exp}.
 Section \ref{sec:analysis} is devoted to the analysis of the experimental data.
 We show that varying the ratio of herders led to qualitative changes in the 
 asymptotic behavior of the convergence of the voting rate.
 In section \ref{sec:model}, we introduce a stochastic model which simulates the system. 
 We obtain a microscopic rule regarding how herders copy others. 
 We performed a simulation study of a stochastic model for $10^{6}$ subjects based on 
 the herder's microscopic rule.  
 The model showed the information cascade transition to the two-peaks phase, where the convergence 
 completely terminates  as the ratio exceeds some critical value.
 Section \ref{sec:conclusions} is devoted to the conclusions. 
 In the appendices, we explain the experimental setup in detail.  

\section{\label{sec:exp}Experimental design and procedure}

\begin{table*}[htbp]
\caption{\label{tab:design}
Experimental design.
}
\begin{tabular}{lccccc}
\hline
Experiment & $T$ & $\{r\}$ & $M$ & Subject pool  & System \\ 
\hline
2010A & 31 & $\{0,1,2,3,5,7,9,\infty\}$ & 100  & Kitasato Univ.    &  Face-to-Face  \\
2010B & 31 & $\{0,1,2,3,5,7,9,\infty\}$ & 100  & Kitasato Univ.    &  Face-to-Face  \\
2011A & 52 & $\{0,1,5,11,21,\infty\}$ & 120  & Hokkaido Univ.    &  Web  \\
2011B & 52 & $\{0,1,5,11,21,\infty\}$ & 120  & Hokkaido Univ.    &  Web  \\
\hline
\end{tabular}
\end{table*}

 The experiments reported here were conducted at the Information Science Laboratory 
 at Kitasato University in October 2010 and at the Group Experiment Laboratory of the 
 Center for Experimental Research in Social Sciences at Hokkaido University between 
 June 2011 and July 2011. The subjects included students from the two universities. 
 We call the former experiment EXP2010 and the latter EXP2011. In EXP2010 (EXP2011), the number 
 of individuals $T$ was 31 (52). We prepared two groups of subjects,  
 group A and group B. In Total, 62 (104) subjects participated in EXP2010 (EXP2011). 
 There were  two sequences of 
 subjects, and we denote the order of each subject by $t\in \{1,2,\cdots,T\}$.
 The number of questions $M$ in the two-choice quiz is 
 100 (120) in EXP2010 (EXP2011). 
 Interaction between subjects in each group was permitted only through the social information given by 
 the experimenter (Face-to-Face) in EXP2010 or by the experiment server (Web) in EXP2011.
 Table \ref{tab:design} summarizes the design.

 The subjects answered the quiz individually with and without 
 information about the previous subjects' choices. This information, called social information, 
 is given as the summary statistics of the previous $r$ 
 subjects $\{C_{0}(r,t),C_{1}(r,t)\}$. 
 We denote the $t+1$-th subject's answer for case $r$ by $X(r,t+1)$, 
 which takes the value 1 (0) if the choice 
 is true (false). $\{C_{0}(r,t),C_{1}(r,t)\}$ are the numbers of subjects who choose each choice among 
 the prior $r$ subjects as $C_{1}(r,t)=\sum_{t'=t-r+1}^{t}X(r,t')$ and $C_{0}(r,t)=r-C_{1}(r,t)$.
 The choice of $r$ is
 $r\in \{r\}=\{0,1,2,3,5,7,9,\infty\}(\{0,1,5,11,21,\infty\})$ in EXP2010 (EXP2011).
 Here, $r=0$ means that the subjects receive no information and must answer with their 
 knowledge only. In the case $r=\infty$, the summary statistic is calculated from all previous 
 subjects' choices.  The subjects answered the quiz 
 with their knowledge only ($r=0$) initially. 
 Next, they answered with $\{C_{0}(r,t),C_{1}(r,t)\}$
 from $r=1$ to $r=\infty$ in increasing order of $r$ in the set $\{r\}$. 
 Any differences between the choices in $r=0$ and $r\ge 1$
 can be attributed to the social information.

\begin{figure}[htbp]
\begin{tabular}{c}
\includegraphics[width=8cm]{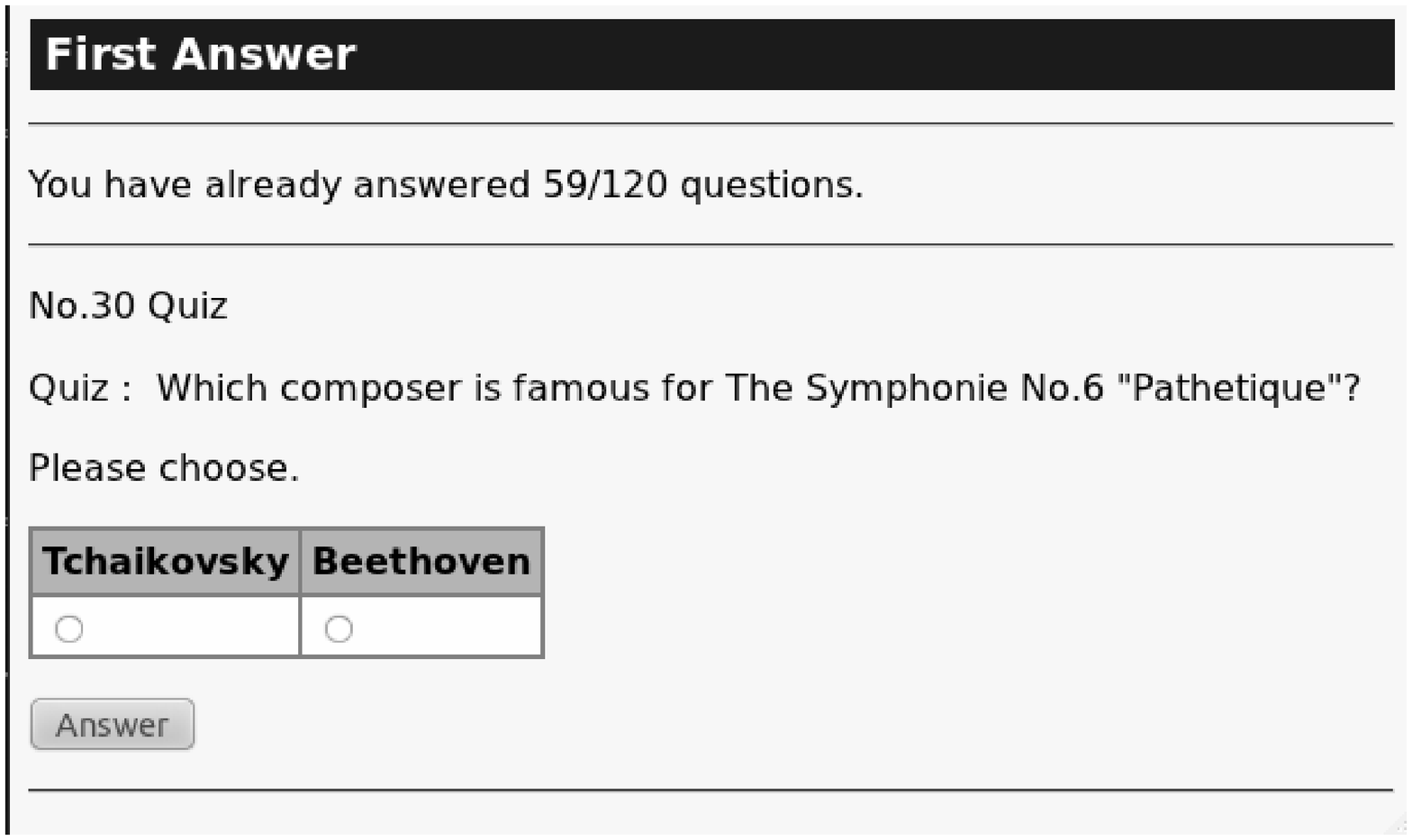} 
\vspace*{0.3cm} \\
\includegraphics[width=8cm]{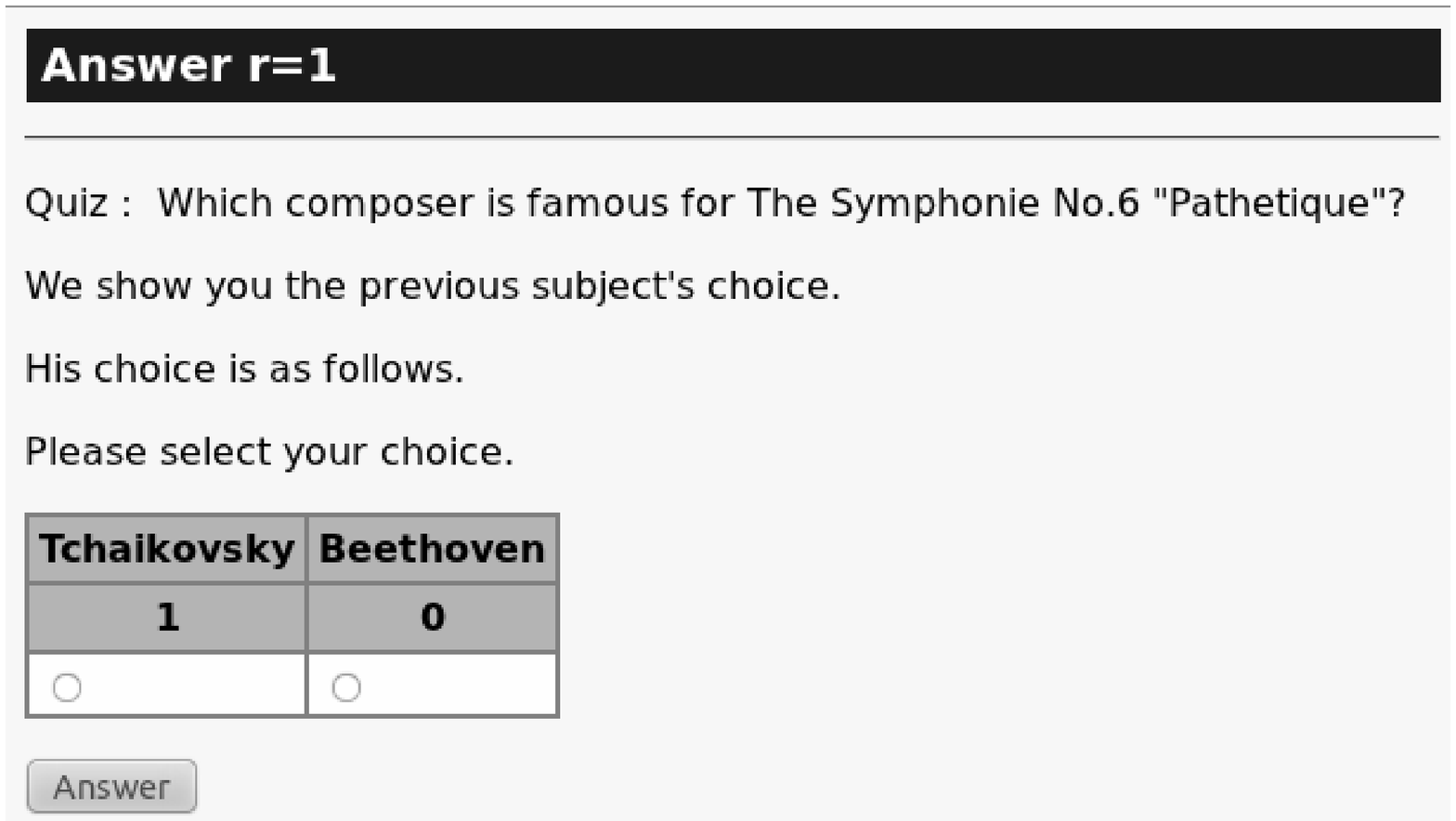} 
\vspace*{0.3cm} \\
\includegraphics[width=8cm]{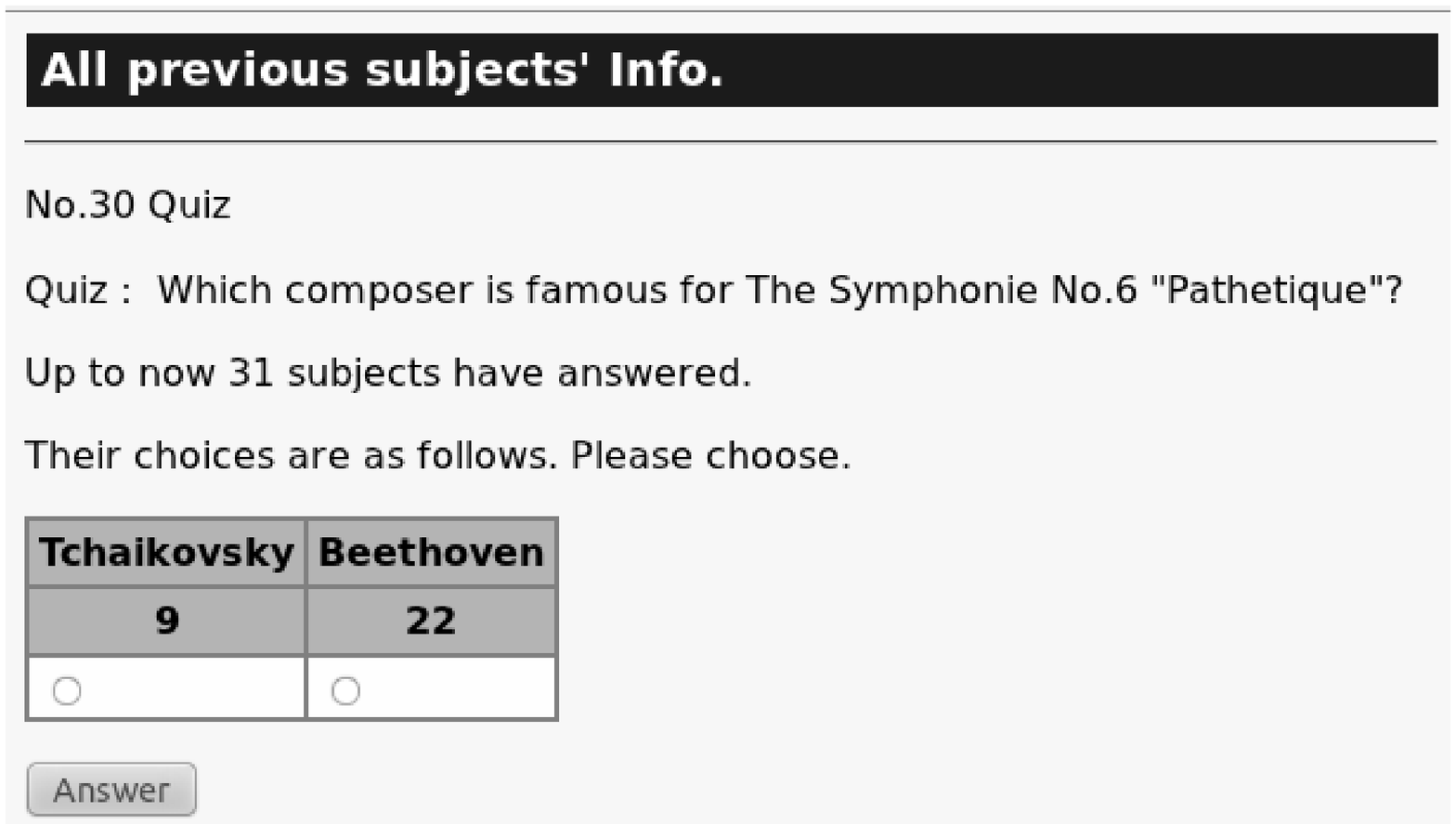} \\
\end{tabular}
\caption{\label{fig:experience} 
Snapshot of the screen for 
$r=0,1,$ and  $\infty$ in EXP2011. 
The summary statistics $\{C_{0}(\infty,t),C_{1}(\infty,t)\}$ are given in the 
second row in the box in cases $r=1$ and $\infty$.}
\end{figure}	

  Fig. \ref{fig:experience} shows the experience of the subjects in EXP2011 more concretely.
  The subjects entered the laboratory and sat in the partitioned spaces.
  After listening to a brief explanation about the experiment and the reward, 
  they logged into the experiment web site using their IDs and started to answer the questions.
  A question was chosen by the experiment server and displayed on the monitor. 
  First, subjects answered the question using their own knowledge only ($r=0$). 
  Later, subjects received social information and answered the same question.
  Fig. \ref{fig:experience} shows the cases $r=1$ and $r=\infty$.
  Subjects could then use or ignore the social information  when making decisions.

\subsection{Experimental procedure}

\begin{figure}[htbp]
\includegraphics[width=9cm]{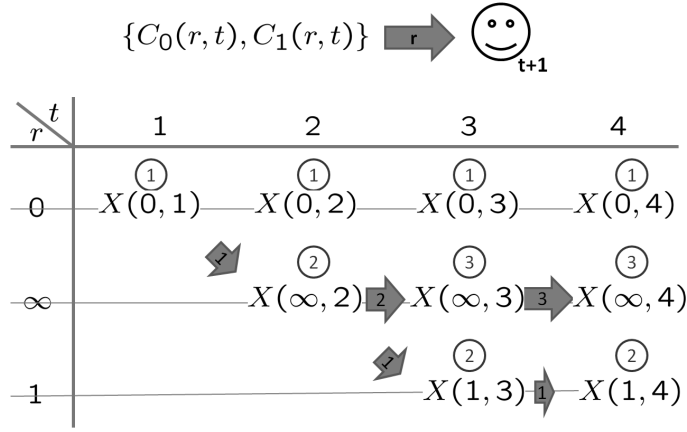}		
\caption{\label{fig:procedure} 
 \color{black}
 Pictorial explanation of experimental procedure. 
 A subject answers the quiz questions in increasing order  of $r$ from $r=0$ to $r=\infty$.
 The $t+1$-th subject answers in case $r$ with social information $\{C_{0}(r,t),C_{1}(r,t)\}$ and the answer 
 is denoted as $X(r,t+1)$. \color{black} 
 The first subject $(t=1)$ answers only in the case $r=0$, which is written as $X(0,1)$.
 The second subject $(t=2)$ answers in the cases $r=0$ and $r=\infty$, which are written as $X(0,2)$ 
 and $X(\infty,2)$, respectively.
 The third subject answers in case $r=0$, then in case $r=1$, and finally in case $r=\infty$; these are 
 written as $X(0,3),X(1,3)$, and $X(\infty,3)$, respectively.
 The number in the circle indicates the order of the answer for each subject. 
 The number in the arrow indicates the memory length of the social information.
 For $r=1$, it is one. For $r=\infty$, the $t$-th subject receives $t-1$ previous subjects' 
 information, and the length is $t-1$.
}
\end{figure}

 We here explain in detail the procedure and  the experience of the subjects in the experiment
\footnote{For more details of the experiments, please refer to Appendix A}.
 A subject answered a question with no public information $(r=0)$ initially.
 The answer was denoted as $X(0,t)$ for the $t$-th subject in the subjects' sequence.
 If the subject was first $(t=1)$, he answered only in the case $r=0$.
 $X(0,1)$ was copied to $X(r,1)$ as $X(r,1)=X(0,1)$ for $r\ge 1$ for later convenience.
 If $t>1$, the experimenter (in EXP2010) or the server (in EXP2011)
 calculated the social information $\{C_{0}(r,t),C_{1}(r,t)\}$ and gave 
 it to the subject. If $t=2$, the subject answered the question in case $r=\infty$ 
 with $\{C_{0}(\infty,1),C_{1}(\infty,1)\}$ and the answer was denoted as $X(\infty,2)$.  
 Here, $C_{1}(\infty,1)$ is $X(0,1)$.
 By the convention $X(r,1)=X(0,1)$ for $r\ge 1$, we can write 
 $C_{1}(\infty,1)=X(\infty,1)$. As $C_{1}(1,1)=C_{1}(\infty,1)$,  
 we copy $X(\infty,2)$ to $X(1,2)$ as $X(1,2)=X(\infty,2)$.
 As in $t=1$, $X(1,2)=X(\infty,2)$ is copied to $X(r,2)$ as $X(r,2)=X(1,2)$ for $r> 1$.  
 If $t=3$, the subject answered the question in case $r=1$ with $\{C_{0}(1,2),C_{1}(1,2)\}$, and the answer was  
 denoted as $X(1,3)$.
 The social information $\{C_{0}(1,2),C_{1}(1,2)\}$ was calculated with the  answer $X(1,2)$ as 
 $C_{1}(1,2)=X(1,2)$ by the copy convention $X(1,2)=X(\infty,2)$. 
 Then, the subject answered in case $r=\infty$ and the answer was denoted as $X(\infty,3)$.
 $C_{1}(\infty,2)$ is  $C_{1}(\infty,2)=X(0,1)+X(1,2)$, which 
 can be written as $C_{1}(\infty,2)=X(\infty,1)+X(\infty,2)$.
 For $r>1$, $X(\infty,3)$ is copied to $X(r,3)$. 
 By the copy convention, the social information in case $r$ can be expressed with 
 the answers $\{X(r,t)\}$ in case $r$ only. 
 If $t=4$, the subject answered the question in case $r=0$, which is written as $X(0,4)$, and  
 then in case $r=1$, written as $X(1,4)$. Next, in EXP2011, the subject  answered in case 
 $r=\infty$, written as $X(\infty,4)$. The social information is $C_{1}(1,3)=X(1,3)$ and 
 $C_{1}(\infty,3)=X(\infty,1)+X(\infty,2)+X(\infty,3)$. 
 Fig. \ref{fig:procedure} gives the pictorial explanation of the procedure. 
 
 In general, if the order of the subject is $t$, there is no 
 public information for $r>t-1$. There is the maximum value $r_{max}$
 in the set $\{r\}$ that satisfies $r_{max}<t-1$.
 The subject answered from case $r=0$ to case $r_{max}$ in the set $\{r\}$ in increasing order of $r$.
 Then, the subject was given 
 the social information from all priors ($r=\infty$) and answered in case $r=\infty$.
 He did not answer cases $r>r_{max}$ in the set $\{r\}$ and 
 the answer in case $r=\infty$ was copied to the unanswered cases as $X(r,t)=X(\infty,t)$ for $r>r_{max}$ in set $\{r\}$.
 The answer in case $r$ started from the $r+2$-th subject in the sequence. 
 For $t<r+2$, $X(r,t)=X(\infty,t)$ by the copy convention, and $\{C_{0}(r,t),C_{1}(r,t)\}$ can be written  
 using only $\{X(r,t)\}$ as $C_{1}(r,t)=\sum_{t'=t-r+1}^{t}X(r,t')$ and $C_{0}(r,t)=r-C_{1}(r,t)$.
 We use the same conventions  when we prepare a sequence of choices in case $r$. 
 All sequences of choices $\{X(r,t)\}$ start from $t=1$.  
 The percentage of correct answers up to the $t$-th subject is defined as  $Z(r,t)=\sum_{t'=1}^{t}X(r,t')/t$, 
  and the final value is  $Z(r,T)$.

\subsection{Quiz selection}

\begin{table*}[htbp]
\caption{\label{tab:questions}
Five typical questions from the  two-choice quiz in EXP2011.
$q$ is the label of the questions in the quiz and $q \in \{0,1,2\cdots,119\}$.}
\begin{tabular}{lcccc}
\hline
q & Question & Choice$_{0}$ & Choice$_{1}$ & Answer \\ 
\hline
0 &  Which insect's wings flap more in one minute? & mosquito & honeybee & 0 \\
1 &  During which period did the Tyrannosaurus Rex live? & Jurassic  & Cretaceous &1 \\
3 &  Which animal has a horn at birth? & rhinoceros & giraffe &1 \\
7 &  Which is forbidden during TV programs in Korea? & commercials & kissing scenes & 0 \\
8 &  Which instrument is in the same group as the marimba ? & vibraphone & xylophone & 1 \\
\hline
\end{tabular}
\end{table*}

  We explain the choice of the questions in the quiz.
  In the experiment, it was necessary to control the difficulty of the questions.
  If a question is too easy, all subjects know the answer.
  If the question is too difficult, no subjects know the answer. 
  In order to study social influence by varying the ratio of people who do not know the answer,
  it is necessary to choose moderately difficult questions.
  In EXP2010, we selected 100 questions 
  for which only one among the five experimenters knew the answer. 
  This choice means that the ratio is estimated to be around $80\%$ for the subjects. 
  After EXP2010, we calculated $Z(0,T)$ for each question.
  In general, $Z(0,T)\ge 0.5$ for two-choice questions.
  A too-small value of $Z(0,T)$ indicates some bias in the given choices of the question.
  We excluded questions with too-small values of $Z(0,T)$  and prepared a new quiz with 120 questions in EXP2011.
  Table \ref{tab:questions} shows five typical questions from EXP2011.

\section{\label{sec:analysis}Data analysis}

 In the analysis, the subjects are classified into two categories, independent and herder, for each question. 
 If a  subject knows the answer to a question with 100\% confidence and 
 the answer is not affected by others' choices, he/she is categorized as independent. 
 If the subject does not know the answer and if he/she may be affected by others' choices,
 he/she is categorized as herder \cite{Mor:2010,His:2010,His:2011}.
 We assume that the probability of a correct choice for independent and herder subjects 
 to be 100\% and 50\%, respectively.
 For a group with $p$ herders and $1-p$  independent subjects, 
 the expectation value of $Z(0,T)$ is $E(Z(0,T))=1-p/2$.
 For each question in each group, we estimate $p$ by $p=2(1-Z(0,T))$ as the maximal likelihood estimate. 
 The assumption of the random guess $(50\%)$ by the herder might be too simple.
 As $Z(0,T)$ approaches $0.5$ and almost all subjects do not know the answer to the question,
 $p$ approaches $100\%$ and the estimate works well. 
 
 Our experimental and analytical design has three advantages 
 over both theoretical models and observational studies.
(i) We control the amount of social information that the subjects receive by the change in  $r$. 
This enables us to derive a 
microscopic rule for human decisions under social information \cite{Lat:1981}. 
(ii) Based on the answers in the absence of information ($r=0$), we can estimate the ratio of herders $p$, which   
 will enable us to extract the herder's decision rule from the results in (i). 
(iii) Our analysis focuses on the asymptotic behavior of the convergence of 
 $Z(r,t)$ with fixed herder's ratio $p$. We clearly see the collective behavior of humans and the qualitative change when $p$ is varied.
In particular, we can study the possibility of the information cascade transition 
\cite{Lee:1993,His:2010,His:2011,Wat:2002,Cur:2006}.

\color{black}
We include a note about the controllability of $p$ in the experiment.
In the experiment, after all $T$ subjects answered, we calculated $Z(0,T)$ and estimated
the herder's ratio $p$ as $p=2(1-Z(0,T))$. It may seem impossible to control $p$ in the experiment, 
but this is not so. We think $p$ is an inherent property of two-choice questions.
If we can estimate $p$ for a large number of  subjects $T>>1$, we can apply the same value 
to the experiments with other groups. We have compared the two values of $Z(0,T)$ of group A and B 
for the same question in EXP2011. Pearson's correlation coefficient $\rho$ is 0.82, and
there is a strong correlation. The system size $T$ in EXP2011 is very limited 
($T\simeq 50$) and there remains some fluctuation in the estimation of $p$, but it will disappear for a 
large enough $T$. We can know $p$ in advance and control it in voting experiments.        
\color{black}

\subsection{Distribution of $Z(r,T)$}

\begin{table}[htbp]
\caption{\label{tab:table}%
Effect of social information on subjects' decisions in (A) EXP2011 and (B) EXP2010.
We divide the samples according to the size of $Z(r,T)$.
$N(r)$ denotes the number of samples in each bin.
$p_{avg}$ is estimated as the average value of $p=2(1-Z(0,T))$ in each bin.  
In the last column, the sub-optimal ratios are shown.   
}
\begin{ruledtabular}
(A) EXP2011
\begin{tabular}{rcccccc}
No. & $Z(r,T)[\%]$ & $N(0)$ & $p_{avg}[\%]$ & $N(5)$ & $N(\infty)$ & Ratio \\ 
\hline
1 &  $<5$ &       0   & NA     &  0  & 2  & NA  \\
2 &  $5\sim 15$ & 0   & NA     &  4  & 18  & NA  \\
3 &  $15\sim 25$& 8   & NA     &  8  & 22 & 8/8  \\
4 &  $25\sim 35$& 16  & NA     &  21  & 20  & 13/16  \\
5 &  $35\sim 45$& 36  & NA     &  24  & 8  & 28/36  \\
6 &  $45\sim 55$& 43  & \color{black}96.7\color{black}   &  26 & 9  & 16/43  \\
7 &  $55\sim 65$& 46  & 79.3   &  34 & 10  & 9/46   \\
8 &  $65\sim 75$& 45  & 62.7   &  40 & 14  & 2/45   \\
9 &  $75\sim 85$& 33  & 41.9   &  45 & 33  & 0/33   \\
10 & $85\sim 95$& 11 &  21.3 &  32 & 67  & 0/11   \\
11 & $\ge 95$    &  2  &  2.0 &  6  & 37  & 0/2   \\
\hline 
Total & & 240 & & 240 & 240  &  76/240 \\ 
\end{tabular}
(B) EXP2010
\begin{tabular}{rcccccc}
No. & $Z(r,31)$ & $N(0)$ & $p_{avg}[\%]$ & $N(5)$ & $N(\infty)$ & Ratio \\ 
\hline
1 & $\le 1/31 $        & 0   & NA    &  0  & 3  & NA  \\
2 & $2/31 \sim 5/31 $  & 2   & NA    &  7  & 20  & 2/2  \\
3 & $6/31 \sim 9/31 $  & 6   & NA    &  18  & 22 & 6/6  \\
4 & $10/31 \sim 13/31 $& 26  & NA    &  21 & 14  & 23/26  \\
5 & $14/31 \sim 17/31 $& 52  & \color{black}97.6\color{black} &  33 & 6  & 26/52  \\
6 & $18/31 \sim 21/31 $& 54  & 74.6  &  21 & 7  & 5/54  \\
7 & $22/31 \sim 25/31 $& 33  & 49.1  &  35 & 29 & 1/33   \\
8 & $26/31 \sim 29/31 $& 25  & 26.3 &  47 & 67  & 0/25   \\
9 & $\ge 30/31 $         & 2   & 6.5   &  18 & 32  & 0/2   \\
\hline 
Total & & 200 & & 200 & 200  &  63/200 \\ 
\end{tabular}
\end{ruledtabular}
\end{table}

There are $2\times 120 (100)$ samples of sequences of choices for each $r$ in EXP2011 (EXP2010).
We divide these samples into 11 (9) bins according to the size 
of $Z(r,T)$, as shown in Table \ref{tab:table}A (\ref{tab:table}B). 
The samples in each bin
share almost the same value of $p$.
For example,
 in the sample in the No. 6 bin ($0.45<Z(0,T)\le 0.55$) in EXP2011, there are almost only 
 herders in the subjects' sequence and $p \simeq 100\%$. 
 On the other hand, in the sample in No. 11 bin ($Z(0,T)>0.95$),
  almost all subjects know the answer to the question and are independent $(p\simeq 0\%)$. 
 An extremely small value of $Z(0,T)$ indicates some bias in the question. 
We omit data with $Z(0,T)<0.45 (13/31)$
 in the analysis of the system  
and we are left with 180 (166) samples in EXP2011 (EXP2010).
\color{black} The samples with $Z(0,T)<0.5$ in No. 6 bin in EXP2011 (No. 5 bin in EXP2010) have $p$ values larger than $100\%$.
These values are errors of the estimation $p=2(1-Z(0,T))$. The standard deviation of $Z(0,T)$ is $p/2\sqrt{T}$ for fixed $p$.
In the estimation of $p$, there is a fluctuation with the magnitude of order $\simeq p/\sqrt{T}$. 
If $p$ takes value larger than $100\%$, we take it to be $100\%$. 
\color{black}
Table \ref{tab:table} shows the number of data samples in each bin for $r=0,5$ and $r=\infty$ as $N(0), N(5),$ and $N(\infty)$. 
Social information causes remarkable  changes in the subjects' choices.
For $r=0$ $(r=5)$, there  is one peak at No. 7 (No. 9), and for $r=\infty$, there are peaks at
No. 3 and No. 11 in EXP2011. \color{black} Here, we compare the densities, not the value $N(r)$ itself. \color{black} 
In the last column, we show the ratio of sub-optimal cases $\{Z(\infty,T)<1/2\}$ with respect to 
the $N(0)$ samples in each bin.
The crucial problem is whether the sub-optimal cases $\{Z(\infty,T)<1/2\}$ 
 remain so in the thermodynamic limit $T\to \infty$ \cite{Lee:1993,His:2011}.

\begin{figure*}[htbp]
\begin{tabular}{ccc}
\resizebox{55mm}{!}{\includegraphics[width=5.5cm]{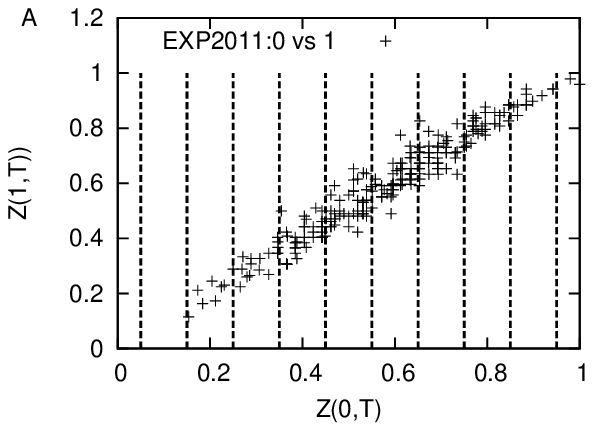}} &
\resizebox{55mm}{!}{\includegraphics[width=5.5cm]{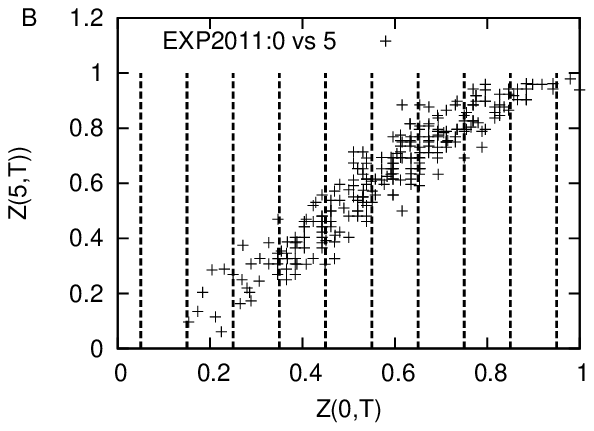}} &
\resizebox{55mm}{!}{\includegraphics[width=5.5cm]{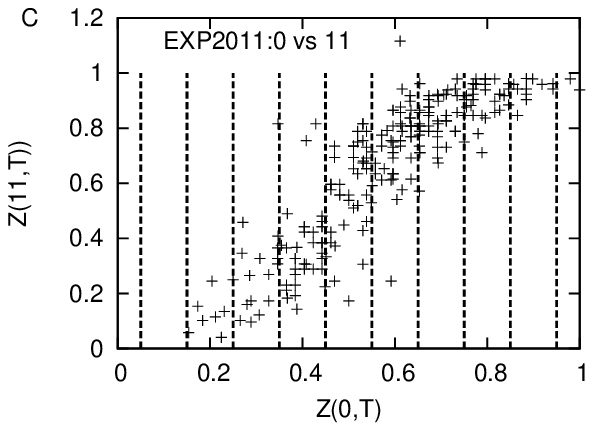}} \\
\resizebox{55mm}{!}{\includegraphics[width=5.5cm]{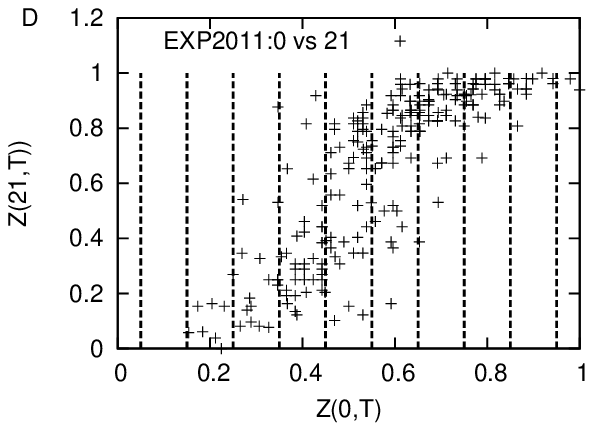}} &
\resizebox{55mm}{!}{\includegraphics[width=5.5cm]{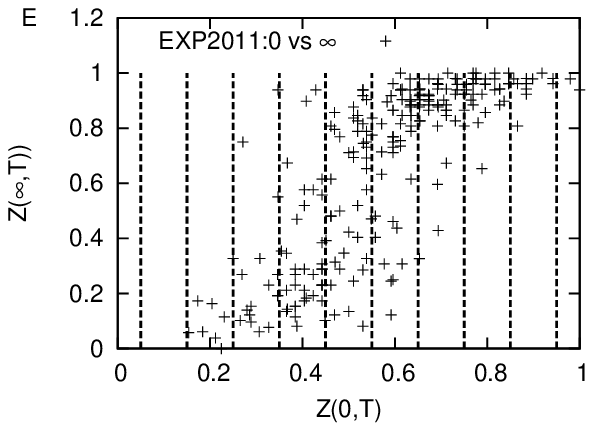}} & \\
\end{tabular}
\caption{\label{fig:scatter_Z}
Scatter plots of $Z(0,T)$ vs $Z(r,T)$ in EXP2011 for (A) $r=1$, (B) $r=5$, (C) $r=11$, (D) $r=21$ and (E) $r=\infty$. 
The vertical lines show the border of the bins
in Table \ref{tab:table}A. 
}
\end{figure*}

 In order to see the social influence more pictorially, we 
 show the scatter plots of $Z(0,T)$ vs $Z(r,T)$ for all 240 samples in EXP2011 
 for each $r\in \{1 (A),5 (B),11 (C),21 (D),\infty (E) \}$ in Fig. \ref{fig:scatter_Z}. 
 The $x$-axis shows $Z(0,T)$ and  the
 y-axis shows $Z(r,T)$. The vertical lines 
 show the boundary between the bins (from No. 1 to No. 11) in Table \ref{tab:table}A. 
 As we move from  Fig. \ref{fig:scatter_Z}A $(r=1)$ to  Fig. \ref{fig:scatter_Z}E $(r=\infty)$, 
 the amount of social information $r$ increases. If the subjects' answers are not affected by the 
 social information, the data should distribute on the diagonal line. However, as the plots clearly indicate,
 this is not the case. As $r$ increases from $r=1$ to $r=\infty$, the changes $Z(r,T)-Z(0,T)$ increase
 and the samples scatter more widely in the plane.
 For the samples with $Z(0,T)\ge 0.75$ (Nos. 9, 10, 11 bins in Table \ref{tab:table}A), 
 the changes are almost positive and $Z(\infty,T)$ takes a value of about one.
 The sub-optimal ratios are zero in the bins. 
 The average performance improves by the social information there. 
 On the other hand, for 
 the samples with $0.45 \le Z(0,T)<0.55$ (No. 6 bin in Table \ref{tab:table}A), 
 the social information does not necessarily improve the 
 average performance. There are many samples with negative change $Z(\infty,T)-Z(0,T)<0$.
 These samples are in the sub-optimal state and constitute the lower  
 peak in Table \ref{tab:table}A.

\subsection{Order parameters of the phase transition}

 We have seen drastic changes in the distribution of $Z(\infty,T)$ from the distribution of $Z(0,T)$.
 Table \ref{tab:table} shows the two-peaks structure in the distribution of $Z(\infty,T)$.  
 In Fig. \ref{fig:scatter_Z}, we see an S-shaped curve in the case $r=\infty$. 
 The natural question is whether these macroscopic changes can be attributed to 
 the information cascade transition. 
 In our previous work on the voting model with digital herders \cite{His:2011}, 
 we showed the possibility of the phase transition from the
 one-peak phase to the two-peaks phase. If $p$ is smaller than some critical value $p_{c}$, the system is in the 
 one-peak phase. The distribution of $Z(\infty,T)$ has only one peak at $Z(\infty,T)>1/2$. 
 If $p>p_{c}$, the distribution of $Z(\infty,T)$ has two peaks at $Z(\infty,T)<1/2$ and $Z(\infty,T)>1/2$.
 In the thermodynamic limit $T\to \infty$, the probability that $Z(\infty,T)<1/2$ becomes a function of $p$, which is non-analytic  
 at $p=p_{c}$ and takes a positive value for $p>p_{c}$.  
 There are two candidates for the  order parameter of the phase transition. One is the sub-optimal ratio $Z(\infty,T)<1/2$. 
 The other is the variance of $Z(\infty,T)$, Var$(Z(\infty,T))$. Both candidates are  zero for $p<p_{c}$ and positive 
 for $p>p_{c}$  in the thermodynamic limit $T\to \infty$.

\begin{figure}[htbp]
\begin{tabular}{c}
\includegraphics[width=8cm]{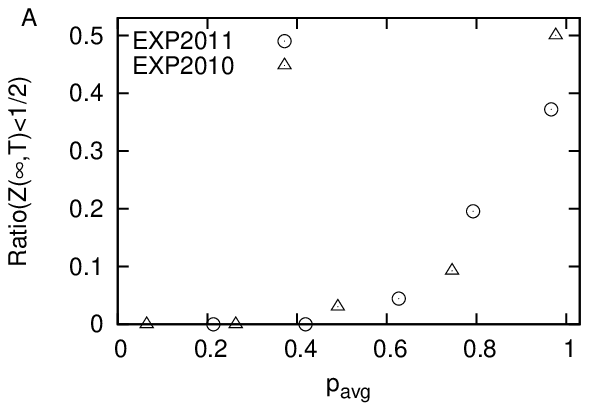} \\
\includegraphics[width=8cm]{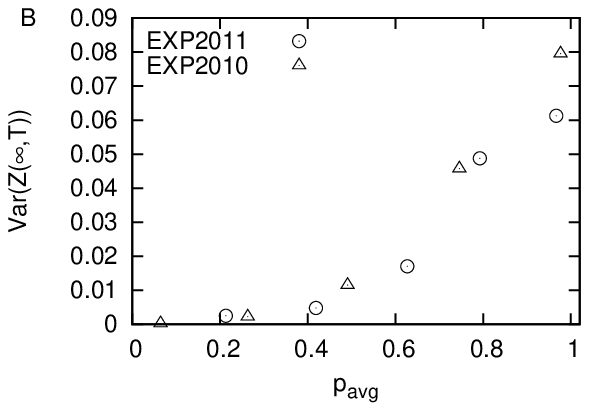} 
\end{tabular}
\caption{\label{fig:p_alpha_var}
(A) The ratios of the sub-optimal $(Z(\infty,T)<1/2)$ cases, which are given in the last column 
 of Table \ref{tab:table}, and (B) the variance Var$(Z(\infty,T))$, are plotted vs $p$.
 The symbols $\circ  (\triangle)$ indicate the results of EXP2011 (EXP2010).}
\end{figure}

Fig. \ref{fig:p_alpha_var} shows the plot of the two order parameters vs $p$.
We plot (A) the ratios of the sub-optimal $(Z(\infty,T)<1/2)$ cases and (B) Var$(Z(\infty,T))$.
The ratios are given in the last columns of Table \ref{tab:table}. 
As $p$ increases, the order parameters change from zero to some finite value.
They are monotonically increasing functions of $p$. However, in the behaviors, we cannot 
 see any clear evidence of the phase transition. The system size is very small and  we cannot see 
 any non-analytic nature there. 
 We cannot use them to prove the existence of the information cascade transition.

\subsection{Asymptotic behavior of the convergence of $Z(\infty,t)$}

We study the convergence of $Z(\infty,t)$ in the limit $t\to \infty$ to clarify
the possibility of  the information cascade transition.
If information aggregation works under social information, 
$Z(\infty,t)$ converges to some value larger than half. 
 The distribution of $Z(r,t)$ has only one peak: it is in the one-peak phase.
 Depending on the convergence behavior, the one-peak phase is classified into two phases.
 If the variance of $Z(r,t)$ shows normal diffusive behavior as  
 $\mbox{Var}((Z(r,t))\propto  t^{-1}$, it is called the normal diffusion phase. 
 We note that the variance is estimated for the ratio, and the usual behavior $t^{1}$ for the sum of 
 $t$ random variables is replaced by $\propto t/t^{2}=t^{-1}$.
 If the convergence is slow and it obeys $\mbox{Var}(Z(r,t))\propto  t^{-\gamma}$ with $0<\gamma<1$,
  it is called the super diffusion phase \cite{Hod:2004,His:2010}. 
 If information aggregation does not work and there is a finite 
 probability that $Z(r,t)$ converges to some value less than half, 
 the distribution of $Z(r,t)$ has two peaks \cite{Lee:1993,His:2011}. It is in the two-peaks phase, and  
 $\mbox{Var}(Z(r,t))$ converges to some finite value in the limit $t\to\infty$.

\begin{figure}[htbp]
\begin{tabular}{c}
\includegraphics[width=9cm]{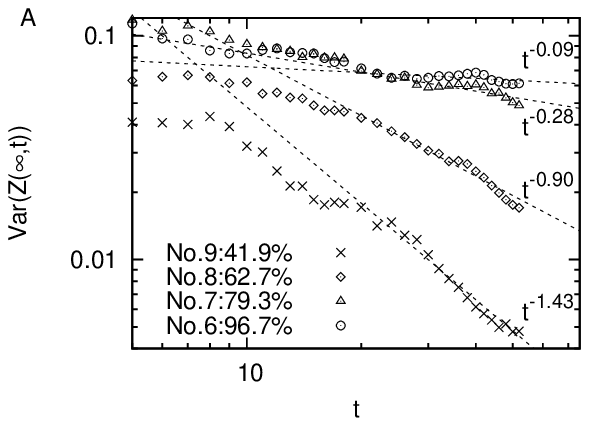} \\
\includegraphics[width=9cm]{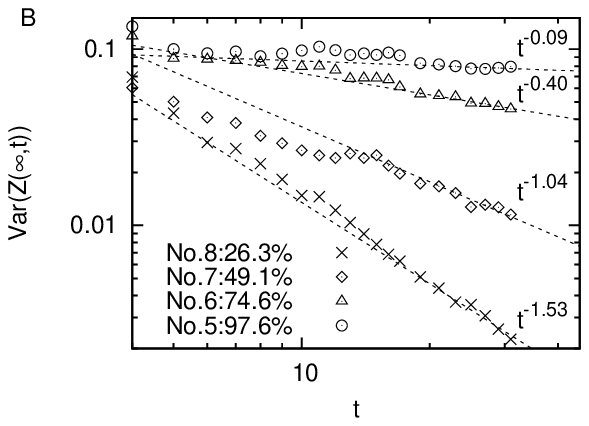} 
\end{tabular}
\caption{\label{fig:herd_macro} 
Convergent behavior for (A) 
EXP2011 and (B) EXP2010.
The convergence is given by the  double logarithmic plot of 
$\mbox{Var}(Z(\infty,t))$ vs $t$ using the samples in the four bins for $r=0$
in Table \ref{tab:table}A (No.6$(\circ)$,7$(\triangle)$,8$(\diamond)$ and 9$(\times)$) 
and in Table \ref{tab:table}B (No.5$(\circ)$,6$(\triangle)$,7$(\diamond)$ and 8$(\times)$).
The dotted lines are fitted results with $\propto t^{-\gamma}$ for $t\ge 20$. 
}
\end{figure}

 Fig. \ref{fig:herd_macro}A (B) shows the double logarithmic plots of $\mbox{Var}(Z(\infty,t))$ as a function of 
 $t$ for EXP2011 (EXP2010).
 If the  plot of Var$(Z(\infty,t))$ vs $t$ has a negative slope $(\gamma>0)$ in the limit $t\to \infty$, the system 
 is in the one-peak phase. If the slope is zero $(\gamma\le 0)$ 
 in the limit  $t\to \infty$, the system is in the two-peaks phase. 
 We see that the convergence becomes very slow as $p$ increases.
 The exponent $\gamma$ is estimated by fitting 
 with $\propto t^{-\gamma}$ for $t\ge 20$.
 It decreases from  $1.43$ $(1.53)$ to $0.09$ with the increase in $p$ in EXP2011 (EXP2010).
 For the cases with \color{black}$p=96.7\% (97.6\%)$\color{black} in EXP2011 (EXP2010), the system
 can be in the two-peaks phase.

\section{\label{sec:model}Stochastic model and Simulation Study}

The asymptotic analysis of the convergence of $Z(\infty,t)$ shows the possibility of the two-peaks phase
 in the cases \color{black}$p=96.7\%$ and $97.6\%$\color{black}. The negative slope $\gamma$ is remarkably small $(\sim 0.09)$ in both experiments.
However, the system sizes are limited and far from the thermodynamic limit $T\to \infty$.
In this section, we derive a microscopic rule as to how the herders copy others' information.
Based on the herder's microscopic rule, we introduce an ad hoc stochastic model.
A simulation study of the model for $10^{6}$ subjects showed the information cascade transition.

\subsection{Microscopic behavior of herders}

We determine how a herder's decision depends on social information. 
For this purpose, we need to subtract the independent subjects'  contribution from $X(r,t+1)$.
The probability of being independent is $1-p$ 
and such a subject always chooses 1 
; the herder's decision 
is then \color{black} simply \color{black}
estimated as 
\footnote{\color{black}More precisely, the probability that $t$-th subject
is independent depends on the choice $X(0,t)$. If $X(0,t)=0$, he is not independent and the probability is zero. 
On the other hand, if $X(0,t)=1$, the probability is $(1-p)/(1-0.5p)$. 
The subtraction should take into account the value of $X(0,t)$. 
Here, we adopt the simple procedure described in the text.\color{black}} 
\[
(X(r,t+1)-(1-p)\cdot 1)/p. 
\] 
The expectation value  of this under $C_{1}(r,t)=n_{1}$
 indicates the probability that a herder chooses an option  under the influence of 
the prior $n_{1}$ subjects among $r$ choosing it.
We denote it by $q_{h}(r,n_{1})$, and it is defined for $t\ge r$ as 
\begin{equation}
q_{h}(r,n_{1})\equiv \mbox{E}((X(r,t+1)-(1-p))/p|C_{1}(r,t)=n_{1}).  \label{q_exp}
\end{equation}
The conditional expectation value in Eq.(\ref{q_exp}) 
is estimated using the samples that satisfy $C_{1}(r,t)=n_{1}$ from EXP2011.
From the symmetry between $1\leftrightarrow 0$, we assume $q_{h}(r,n_{1})=1-q_{h}(r,r-n_{1})$. 
For $r=\infty$, the $t+1$-th subject obtains information from 
the previous $t$ subjects and $r$ is considered to be $t$.
\color{black}$r=\infty$ case is averaging several values of $r$. \color{black}
For $r=21$ and $\infty$, we study the dependence of $q_{h}(r,n_{1})$ on
$n_{1}/r$ and round $n_{1}/r$ to the nearest values in $\{k/11 |k\in \{0,1,2,\cdots,11\}\}$. 
In addition, we estimate $q_{h}(\infty,n_{1})$ for $t\ge 22$ to understand the herder's 
decision under the largest amount of social information. 
\color{black}
The second reason is  
that the subjects receives much information for $t\ge 22$ and we can assume that 
the dependence of $q_{h}(\infty,n_{1})$ on $(t,n1)$ is replaced 
with the dependence on $n1/t$.
\color{black}

\begin{figure}[h]
\includegraphics[width=9cm]{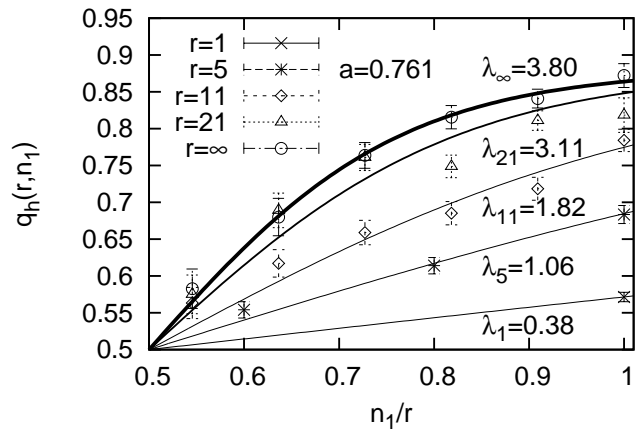}
\caption{\label{fig:herd_micro} 
Microscopic rule of herder's decision in EXP2011. 
It shows the probability $q_{h}(r,n_{1})$ 
that a herder chooses an option under the influence of the prior $n_{1}$ 
subjects among $r$ choosing it
for cases $r=1(\times),5(\ast),11(\diamond)$ and $21(\triangle)$.
For $r=\infty(\circ)$, $r$ is considered to be $t$.
The thin solid curves are fitted results 
with Eq.(\ref{q_model}) for $r=1,5,11,$ and $21$ from the bottom.
The top thick solid line corresponds to $r=\infty$.}
\end{figure}

Figure \ref{fig:herd_micro} shows the plot of $q_{h}(r,n_{1})$ vs $n_{1}/r$.
It is clear that $q_{h}(r,n_{1})$ is an almost  monotonic increasing function of $n_{1}$.
As $r$ increases, it shows stronger dependence on $n_{1}/r$, and  the 
herder's decision is affected more greatly by the prior subjects' choices. 
We fit the plot by the following functional form:
\begin{equation}
q_{h}(r,n_{1})=\frac{1}{2} \left( a \tanh (\lambda_{r}(n_{1}/r-1/2))+1 \right). \label{q_model}
\end{equation}
The parameters $a$ and $\lambda_{r}$ indicate the strength of the conformity of the subjects. 
Social psychology studies suggest that  people's likelihood to use
 social information depends on their mood  \cite{Gri:2006}. $a$ denotes the net ratio of herders who react 
 positively to the priors subjects' choices.
$\lambda_{r}$ denotes the strength of the dependence on social information.
By the least squares fit,  we obtain \color{black} $a=0.761$ and $\lambda_{\infty}=3.80$ \color{black}  for $r=\infty$. 
The fitted result is also shown in Fig. \ref{fig:herd_micro}. 
The values of $a$ and $\lambda_{\infty}$ depend on the experimental situation and on the system size $T$ 
\footnote{The microscopic behavior of the herders in EXP2010 is given in Appendix B.}.
Using the same $a$, we fit the data for other $r$ using  Eq.(\ref{q_model}). The results are also given 
in Fig. \ref{fig:herd_micro}. 
As the amount $r$ of social information increases, the strength of dependence $\lambda_{r}$ increases.

\subsection{Information cascade transition of voting model}

To understand the behavior of the system in the thermodynamic limit $T\to \infty$, 
we simulate the system for large $T$
by  a stochastic model based  on Eq.(\ref{q_model}). 
We introduce a stochastic process $\{X(t)\},t\in \{1,2,3,\cdots,T\}$.
$X(t+1) \in \{0,1\}$  is a Bernoulli random variable and its probabilistic rule depends on all
the previous $\{X(t')\},t'\in \{1,\cdots,t\}$ through $C_{1}(t)=\sum_{t'=1}^{t}X(t')$.  
The probability that $X(t+1)$ is 1 for $C_{1}(t)=n_{1}$, 
which is denoted as $q(t,n_{1})$, is given as 
\begin{equation}
q(t,n_{1})
=(1-p)+p \cdot \frac{1}{2} \left(a \tanh \lambda_{t}(n_{1}/t-1/2)+1 \right).
\end{equation}
We set $a$ as $a=0.761$ and $\lambda_{t}$ for $t\in \{1,5,11,21\}$ as $\lambda_{t}=\lambda_{r}$.
Here, $\lambda_{r}$ takes the values given in Fig. \ref{fig:herd_micro}. 
For other values of $t$ and $t< 21$, we use the linearly extrapolated value. 
For $t>21$, we set $\lambda_{t}=\lambda_{\infty}$.

We denote the  probability function Pr($\sum_{t'=1}^{t}X(t')=n)$ as $P(t,n)$. 
The  master equation for $P(t,n)$ is
\begin{equation}
P(t+1,n)=q(t,n-1) \cdot P(t,n-1)+(1-q(t,n))\cdot P(t,n).
\end{equation}
We solved the master equation recursively and obtain $P(t,n)$ for $t\le 10^{6}$.

\begin{figure}
\begin{tabular}{c}
\includegraphics[width=8cm]{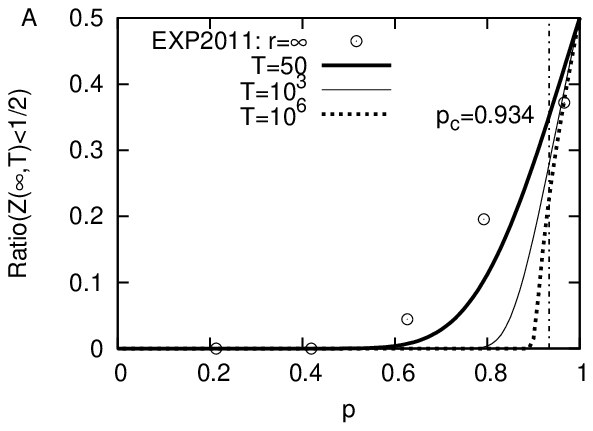} \\
\includegraphics[width=8cm]{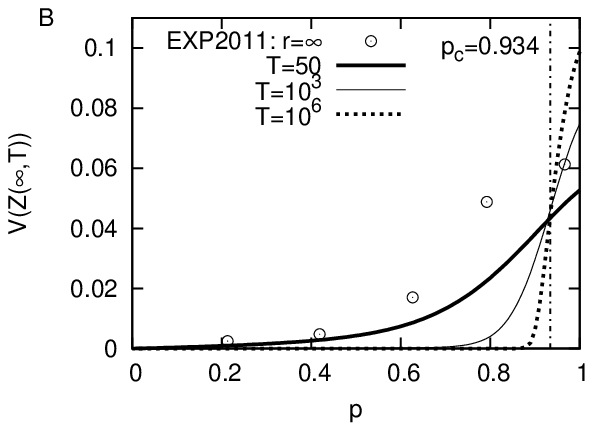}
\end{tabular}
\caption{\label{fig:p_alpha_var_model}
Plots of order parameters vs $p$ for the voting model and the limit $T\to \infty$. 
We plot (A) the sub-optimal ratio $Z(\infty,T)<1/2$ and (B) $\mbox{Var}(Z(\infty,T))$ 
vs $p$; here, $T$ is the length of the sequence. 
The symbol ($\circ$) indicates the experimental data for EXP2011$ (r=\infty)$
($T\simeq 50$) for the five bins No. 6, 7, 8, 9, and 10 in Table \ref{tab:table}.
The lines show the data from the stochastic model for $r=\infty$. 
For the stochastic model, we set 
$T=50$ (thick solid line), $10^{3}$ (thin solid), and $10^{6}$ (thick dotted).
As $T$ increases, the non-zero regions of the order parameters move rightward.
In the limit $T\to \infty$, the region reduces to \color{black}$p>p_{c}=93.4\%$.\color{black}
The vertical chain line at $p=p_{c}$ shows the critical point of 
the phase transition between the one-peak phase $(p<p_{c})$ 
and the two-peaks phase $(p>p_{c})$.
}
\end{figure}

Fig. \ref{fig:p_alpha_var_model} shows the results of 
the model. We plot (A) the sub-optimal ratio $Z(\infty,T)<1/2$ and 
(B) $\mbox{Var}(Z(\infty,T))$ vs $p$. 
For comparison, we plot the experimental results (EXP2011) using the symbol $\circ$. 
The model with $T=50$ well describes the experimental results quantitatively.
In the figure, we also plot the results for $T=10^{3}$ and $10^{6}$. 
As $T$ increases, the non-zero regions of the order parameters move rightward. 
In particular, we see that there is a crossing point in the curves of Var$(Z(\infty,T))$ at \color{black}$p=93.4\%$\color{black}, 
 which we denote as $p_{c}$.
If $p<p_{c}$, $\mbox{Var}(Z(\infty,T))$ 
goes to zero as $T$ increases. On the other hand, if $p>p_{c}$, the variance seems to remain in the limit $T\to \infty$. 
This shows the phase transition between the one-peak and  the two-peaks phases. 
For $p>p_{c}$, the system is in the two-peaks phase ($\gamma=0$). 
For $p<p_{c}$, the system is in the one-peak phase ($0<\gamma\le 1$). 
In order to see the convergence rate (the exponent $\gamma$), 
 it is necessary to study the asymptotic behavior of 
 $\mbox{Var}(Z(\infty,t))$ for large values of $t$.

\begin{figure}[h]
\includegraphics[width=9cm]{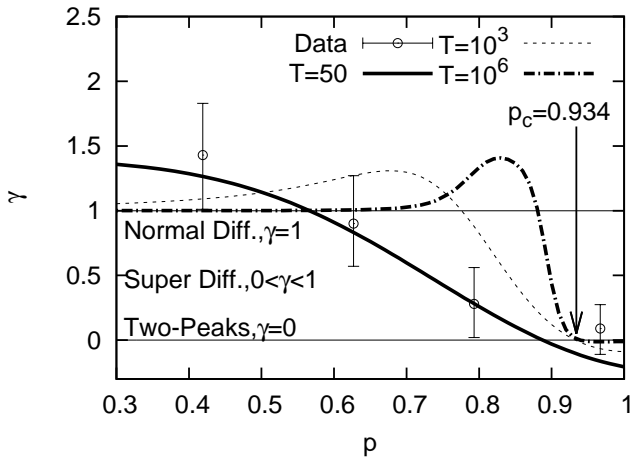}
\caption{\label{fig:herd_model} 
Thermodynamic limit and phase diagram.
Asymptotic behavior of the convergence is given by the plot of $\gamma$ vs $p$. 
The symbols ($\circ$) show the $\gamma$s vs $p_{avg}$ in Fig. \ref{fig:herd_macro}A.
The lines show the 
results of the stochastic model with the system size $T=50$ (thick solid), 
$10^{3}$ (dotted), and $10^{6}$ (thick chain).
}
\end{figure}

We estimate $\gamma$ from the slope
of $\mbox{Var}(Z(t))$ as
\begin{equation}
\gamma=\log \frac{\mbox{Var}(Z(T-\Delta T))}{\mbox{Var}(Z(T))}/\log \frac{T}{T-\Delta T}
\end{equation}
for the time horizons $T=50,10^{3}$ and $10^{6}$. 
For $T=50$, we take $\Delta T=30$ to match
the analysis of the experimental data in EXP2011. For $T=10^{3}$ and $10^{6}$, we take $\Delta T=10^{2}$.    
The results are summarized in Fig. \ref{fig:herd_model}. 
For $T=50$, $\gamma$ is a monotonic decreasing function of $p$ and it describes 
the experimental results of EXP2011 well.
For the limit $T\to \infty$, we compare the results 
with $T=50,10^{3}$, and $10^{6}$. $\gamma$ shows non-monotonic behavior as a function of 
$p$ for the latter two cases; it is an artifact of finite $T$.  In the limit $T\to \infty$, $\gamma$ 
monotonically decreases from 1 to 0, and the threshold value is \color{black}$p_{c}=93.4\%$\color{black} \cite{His:pre}. 
For $p<p_{c}$ $(p>p_{c})$, the system is in the  one-peak (two-peaks) phase.

\section{\label{sec:conclusions}Conclusions}

The instinct to imitate others led to the remarkably slow  convergence of information aggregation 
as the herder's ratio $p$ approached $100\%$. 
A stochastic model based on the herder's microscopic copying rule predicted the information cascade 
 transition between the one-peak ($p<p_{c}$) and  
the two-peaks phases ($p>p_{c}$) \cite{His:pre}.
In the one-peak phase, information aggregation works and the majority's choice is always correct when 
the independents choose the correct choice rather than the incorrect one.
In the two-peaks phase, the majority's choice is not necessarily correct.
The coexistence of the optimal $(Z(\infty,T)>1/2)$ and the sub-optimal $(Z(\infty,T)<1/2)$ states
occurs there. 

It has been thought that information cascade was fragile \cite{Dev:1996,And:1997} or self-correcting
where the sub-optimal state disappears and switches to the optimal state \cite{Goe:2007}. 
Our study indicates that the system is in the two-peaks phase 
and that the sub-optimal state is stable against small perturbations for $p>p_{c}$ if 
 the subjects are given the summary statistics $\{C_{0}(\infty,t),C_{1}(\infty,t)\}$. \color{black}
The conclusion might appear to be contradictory to the previous one, but this is not so.
In the previous works, each subject has his own information and it is not necessary 
to follow the majority if one can trust his own information \cite{Goe:2007}.
In addition, the social information is the time series of the previous choices
$\{X(\infty,t')\}_{1\le t' \le t}$, and it contains much more information than the summary statistics 
$\{C_{0}(\infty,t),C_{1}(\infty,t)\}$. In our experiment, the herder does not have information 
and it is necessary to follow the majority if he wants to choose a correct answer.
\color{black}
However, the system size in our experiment is very limited and it is difficult to infer the state of the system 
in the thermodynamic limit  based only on experimental data.
Our conclusion that there occurs a information cascade transition relies heavily on the results of
the simulation study of the  stochastic model. 
In addition, the experiments were performed with  
students at universities, and the scope of the subjects is thus very restricted. 
The robustness of the conclusion should be established by further experiments.
\color{black}
For this purpose, a web-based experiment in artificial laboratories is promising \cite{Sal:2006}.
There, we can approach the thermodynamic limit $T\to \infty$ more easily than in physical laboratories
 and study the micro-macro feature of information cascade. 
\color{black}

\begin{acknowledgments}
We thank Yosuke Irie  for preparing the quiz used in the experiment 
and Fumihiko Nakamura and Ruokang Han for their assistance in recruiting the subjects.
This work was supported by Grant-in-Aid for Challenging Exploratory Research 21654054.
\end{acknowledgments}

\bibliography{paper_v4}

\appendix

\section{Experimental Setup}

\subsection{EXP2010}

 In EXP2010, the 62 subjects who participated in the experiment were recruited from 
 the School of Science of Kitasato University. 
 The subjects were randomly assigned to either group A or group B; each group had 31 subjects.
 In each session, one subject from each group entered the laboratory, a total of two subjects.
 We explained that we were studying how their choices were affected by the choices of others.
 After explaining the details of the experimental procedure and payment, 
 each subject sat in front of an experimenter and had no contact with the other subject in the laboratory.
 The experiments on groups A and B were performed independently. 
 Interaction between subjects in each group was permitted only through the social information given by 
 the experimenter in front of each subject.
 The subjects were asked to answer the 100 questions in the two-choice quiz.
 The subjects answered the quiz with at most eight social influence conditions, including the case $r=0$.
 Each session lasted about one hour.
 In order to obtain data from all $T=31$ subjects in both groups, 
 we performed the session 11 times on October 9 and 
 and 10 times on October 16 and  23 in 2010.
 Subjects were paid upon being released from the session.  
 There was a 2000 yen (about \$24) participation fee and an additional 1000-yen reward (about \$12) 
 for the top ten subjects.
 The ranking of the subjects was calculated based on the ratio of correct answers to all questions and $r$.

\subsection{EXP2011}

 In EXP2011, the 104 subjects who participated were recruited from the Literature 
 Department of Hokkaido University. The subjects were randomly assigned to either group A or group B; 
 each group had 52 subjects. 
 In each session, between one and six subjects from each group entered the laboratory, for a total of between two and  
 eleven subjects.
 The experiment was performed in the Group Experiment Laboratory of  
 the Center for Experimental Research in Social Sciences of Hokkaido University.
 There were fifteen desks furnished with partitions and PCs.
 Each subject sat behind a partition and communication among them was prohibited.     
 The experiments on groups A and B were performed independently. 
 In order to study the effect of social information on the choices of the subjects, 
 it was necessary to control the transmission of information from others in the same group.
 We developed a web-based  voting system by which multiple subjects could simultaneously participate in  
 the experiment.
 The subjects used a web browser to access the web voting server in the intranet.
 They could obtain information about the others' choices from the summary 
 statistics $\{C_{0}(r,t),C_{1}(r,t)\}$ shown on the monitor.
 Subjects did  not know which questions the other subjects were answering or what their choices were. 

 With slides, we showed subjects how the experiment would proceed.
 We explained that we were studying how their choices were affected by the choices of others.
 In particular, we emphasized  that 
 the social information was realistic information calculated from the choices of previous subjects.
 The reason for this is that in contrast to EXP2010, the social information was given on the monitor and 
 seemed less credible to the subjects.   
 Through the slides, we also explained the payment.
 After the explanation, the experiment started.
 The subjects answered the 120-question quiz with at most six social influence 
 conditions within about one hour.
 A total of 15 sessions were held on  June 15 and 16 and from July 12 and 13 in 2011.
 Subjects were paid in cash upon being released from the session.
 There was a 600 yen (about \$7) participation fee and additional rewards that were proportional to the number
 of correct answers. One correct choice was worth one point, and this was worth one yen (about one and a third cents). 

 Some subjects could not answer all the questions within the alloted time, so   
 the number $T$ of subjects who answered a question varied.
 The distribution of $T$ is $\#\{T=52\}=143,\#\{T=51\}=5$,
 $\#\{T=50\}=2,\#\{T=49\}=85,\#\{T=48\}=4,\#\{T=47\}=1$. 
 The average value of $T$ was $50.8$ and the standard deviation $\sigma_{T}$ was $1.48$. 
 The average length of the sequence was about 50.

 The assignment of quiz questions in EXP2011 was as follows. 
 Each question has a label $q$ in $q\in \{0,1,2,\cdots,119\}$ and 
 each subject has an ID number  $i$.
 If the ID number  $i$ of the subject was odd, the subject started with the question with the smallest $q$ in the pool
 not chosen by the server for another subject in the same group. 
 If $i$ was even, the subject started with the question with the largest $q$ not chosen 
 by the server for another subject.  
 If another subject was answering a question, that question was never chosen by the server.
 It was strictly prohibited for multiple subjects to answer the same question simultaneously.

\section{Microscopic behavior of herders in EXP2010}

\begin{figure}
\includegraphics[width=9cm]{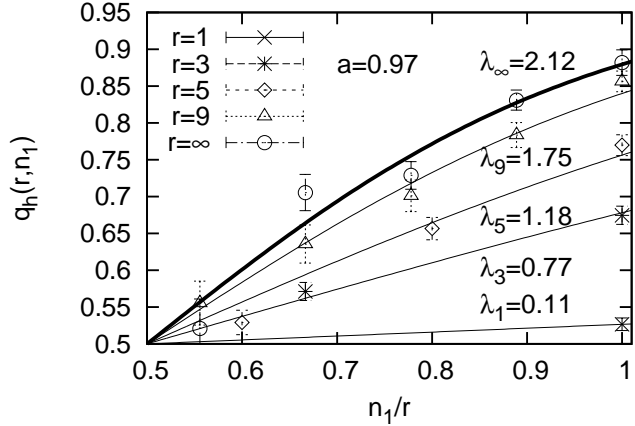}
\caption{\label{fig:herd_micro_2010}
Microscopic rule of herders' decisions in EXP2010. 
It shows the probability $q_{h}(r,n_{1})$ 
that a herder chooses an option under the influence of the prior $n_{1}$ 
subjects among $r$ choosing it
for cases $r=1(\times),3(\ast),5(\diamond)$ and $9(\triangle)$.
For $r=\infty(\circ)$, the $t+1$-th subject's decision depends
on previous $t$ subjects' decisions and $r$ is considered to be $t$.
We estimate the dependences of $q_{h}(\infty,n_{1})$ on
$n_{1}/t$ for $t\ge 10$, where $n_{1}/t$ is rounded 
to the nearest value in $\{k/9 |k\in \{0,1,2,\cdots,9\}\}$. 
The solid curves are fitted results with $q_{h}(r,n_{1})=(a \tanh (\lambda_{r}(n_{1}/r-1/2))+1)/2$.
For $r=\infty$ (the thick solid line), we set $r=t$. 
From bottom to top, the thin solid lines correspond to $r=1,3,5,$ and $9$. 
}
\end{figure}

We determine how a herder's decision depends on social information in EXP2010. 
We follow the same procedure written in the main text. 
Fig. \ref{fig:herd_micro_2010} plots the results.
The fitted results with Eq.(\ref{q_model}) are also shown. 
Compared with Fig. \ref{fig:herd_micro}, the convergence to Eq.(\ref{q_model})
is not so good. The system size $T=31$ is not enough to derive the microscopic rule. 

\section{Estimate of $\gamma$ and its error bar}

 In our study, we focus on the asymptotic behavior of $\mbox{Var}(Z(\infty,t))$. 
 In particular, we are interested in the power law behavior as
 $\mbox{Var}(Z(\infty,t)) \propto t^{-\gamma}$. 
 The negative slope $-\gamma$ of the double logarithmic plot of $\mbox{Var}(Z(\infty,t)$ vs $t$ gives 
 the exponent $\gamma$ in the limit $t\to \infty$.
 In our analysis, we estimate $\gamma$ by the least squares fit with the functional form $a\cdot t^{-\gamma}$ in the 
 range $20\le t \le T$. We denote the estimate as $\gamma_{exp}$.

 For  the error bar of the exponent $\gamma_{exp}$, 
 we adopted the voting model to simulate the system and 
 apply the parametric bootstrapping method based on it.
 First, we solved the model recursively up to $T=50-30$ and $T=50$ and obtained the probability
 functions $P(T,n)$ for both time horizons $T$. 
 We defined $Z(T)$ as $Z(T)=\frac{n}{T}$ and estimated $\gamma$ by the relation 
\begin{equation}
\gamma=\log \frac{\mbox{Var}(Z(T-\Delta T))}{\mbox{Var}(Z(T))}/\log \frac{T}{T-\Delta T} 
\end{equation}
 using the probability function $P(T,n)$.
Here, $\gamma$ was estimated in the range $T-\Delta T \le t \le T$. 
For $T=50$, we take $\Delta T=30$ to match
the analysis of the experimental data in EXP2011. 
The estimate is exact, and we denote it as $\gamma_{exact}$.
In EXP2011, the number of samples in each bin 
ranges from 33 to 46, and is very limited.
We studied the stochastic model using the  Monte Carlo method 
with the same sample size as the experimental data in each bin
 and estimated $\gamma$ for the samples.
We repeated this $10^{4}$ times to obtain the samples of $\gamma$. 
Using the samples of $\gamma$, we estimated 
the 95\% confidence interval, which is denoted as $[\gamma_{-},\gamma_{+}]$. 
The approximately estimated $\gamma$ was distributed around $\gamma_{exact}$. 
The upper (lower) deviation was calculated as $\Delta \gamma_{+}\equiv \gamma_{+}-\gamma_{exact}$ ($\Delta \gamma_{-} \equiv \gamma_{exact}-\gamma_{-}$). 
We estimated the 95\% confidence interval of $\gamma_{exp}$ as
$[\gamma_{exp}-\Delta \gamma_{-},\gamma_{exp}+\Delta \gamma_{+}]$. 
Using this procedure, we estimated the error bars for each $\gamma$ in Fig. \ref{fig:herd_model}.

\end{document}